# Mobility Stream Processing on NebulaStream and MEOS

Mariana M. Garcez Duarte
Université libre de Bruxelles
Brussels, Belgium
mariana.machado.garcez.duarte@ulb.be

Dwi P. A. Nugroho
BIFOLD, TU-Berlin
Berlin, Germany
d.nugroho@tu-berlin.de

Georges Tod
SNCB-NMBS Engineering
Brussels, Belgium
georges.tod@belgiantrain.be

Evert Bevernage
SNCB-NMBS Engineering
Brussels, Belgium
evert.bevernage@belgiantrain.be

Pieter Moelans
SNCB-NMBS Engineering
Brussels, Belgium
pieter.moelans@belgiantrain.be

Emine Tas
SNCB-NMBS Engineering
Brussels, Belgium
emine.tas@belgiantrain.be

Esteban Zimányi
Université libre de Bruxelles
Brussels, Belgium
esteban.zimanyi@ulb.be

Mahmoud Sakr
Université libre de Bruxelles
Brussels, Belgium
mahmoud.sakr@ulb.be

Steffen Zeuch
BIFOLD, TU-Berlin
Berlin, Germany
steffen.zeuch@tu-berlin.de

Volker Markl
BIFOLD, TU-Berlin, DFKI
Berlin, Germany
volker.markl@tu-berlin.de

## Abstract

The increasing use of Internet-of-Things (IoT) sensors in moving objects has resulted in vast amounts of spatiotemporal streaming data. To analyze this data in situ, real-time spatiotemporal processing is needed. However, current stream processing systems designed for IoT environments often lack spatiotemporal processing capabilities, and existing spatiotemporal libraries primarily focus on analyzing historical data. This gap makes performing real-time spatiotemporal analytics challenging.

In this demonstration, we present NebulaMEOS, which combines MEOS (Mobility Engine Open Source), a spatiotemporal processing library, with NebulaStream, a scalable data management system for IoT applications. By integrating MEOS into NebulaStream, NebulaMEOS utilizes spatiotemporal functionalities to process and analyze streaming data in real-time. We demonstrate NebulaMEOS by querying data streamed from edge devices on trains by the Société Nationale des Chemins de fer Belges (SNCB). Visitors can experience demonstrations of geofencing and geospatial complex event processing, visualizing real-time train operations and environmental impacts.

## 1 Introduction

Over the last decade, the use of Internet-of-Things (IoT) enabled new applications for domains such as public transport management [3], urban planning, and maritime traffic management. These applications are supported by the ability to collect and transmit continuous spatiotemporal data streams from IoT sensors embedded in mobile and stationary infrastructure. Nonetheless, while this data enables real-time decision-making, the challenge lies in processing it efficiently and effectively in highly distributed IoT environments. Existing real-time processing solutions with the scalability needed for dynamic, data-intensive applications often lack spatiotemporal awareness for processing such spatiotemporal events, leaving a critical gap in existing IoT systems. Consequently, there is a need for an IoT data management platform that supports low-latency spatial analytics [9]. Moreover, addressing the scalability concern in highly distributed IoT setups requires specialized spatiotemporal algorithms that run on resource-constrained edge devices.

In this work, we present NebulaMEOS, which combines NebulaStream [8] and MEOS [11, 12]. NebulaStream is an end-to-end data management system designed for large-scale IoT use cases. NebulaStream processes sensor data from many devices and supports flexible deployment, even on resource-limited edge hardware. MEOS is an open-source library that manages temporal and spatiotemporal data with efficient data structures, such as spatiotemporal sequences and bounding boxes. Its optimized implementation allows MEOS to run on low-end edge devices, such as a Raspberry Pi, making it ideal for environments with strict hardware constraints.

NebulaMEOS extends NebulaStream with spatiotemporal processing capabilities, addressing the lack of native spatiotemporal support in many streaming systems. While Kafka [6] and Flink [1] are popular for large-scale data streaming, they do not natively manage spatiotemporal analytics [7]. Users must create custom code or integrate separate frameworks, which can lead to complexity and resource overhead. However, integrating spatiotemporal operations into existing streaming infrastructures remains challenging, primarily due to the need for domain-specific extensions. In contrast, NebulaStream offers a more unified and lightweight plug-in mechanism [4] suitable for seamless and reliable integration. Real-time spatiotemporal processing must be both low-latency and workload-adaptive, adjusting to data volume and rate oscillations to maintain consistent throughput in dynamic IoT environments.

NebulaMEOS provides real-time spatiotemporal imputation and analytics in IoT environments, addressing distributed data management challenges and a lack of native spatiotemporal capabilities in existing stream processing frameworks. We demonstrate this integration with a practical example using data from the Société Nationale des Chemins de fer Belges (SNCB), Belgium's national railway operator. SNCB installed sensors and edge devices aboard six trains to monitor vehicle metrics. Currently, these devices send raw data to the cloud without onboard real-time processing. NebulaMEOS enables local, real-time analytics, even when connectivity and hardware resources are limited. This demo delivers the following contributions:

(1) **Real-Time Spatiotemporal Processing:** The demo shows that NebulaMEOS can perform real-time spatiotemporal analytics on distributed IoT networks.
(2) **Scalable and Extensible Integration:** The demo denotes the extensibility of NebulaStream through the incorporation of MEOS operations.
(3) **Showcase of Practical Use Cases:** We demonstrate geofencing and geospatial complex event processing (GCEP) queries. These applications are relevant for railway operators, such as SNCB, which faces related challenges in managing train operations. For example, geofencing can improve safety by ensuring speed limits to maintenance zones, and GCEP enables proactive monitoring of train battery levels.

## 2 System Overview

Our demonstration integrates NebulaStream and MEOS (Mobility Engine Open Source) to manage and process spatiotemporal data in highly dynamic and distributed Internet-of-Things (IoT) environments. By leveraging the strengths of both systems, NebulaMEOS provides a scalable framework for data handling and query execution in scenarios involving mobility data.

Figure 1 shows the architecture of NebulaMEOS. The train has a variety of sensors that provide readings such as brake pressure, speed, temperature, battery status, and geographic location, which are processed by the system to monitor and optimize train operations. These sensors send their data to a centralized processor edge device with an Intel Atom Dual-Core onboard the train. This edge device, which incorporates the proposed system, processes the incoming sensor data and transmits the processed data to a server for visualization using Deck.gl.[1] The query processing workflow uses NebulaStream's distributed capabilities. It employs its coordinators and worker nodes to manage computations and allows execution directly on edge devices.

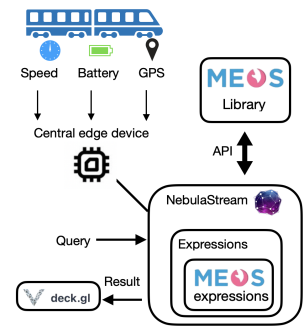

Figure 1: System Overview

### 2.1 NebulaStream

NebulaStream [8] is an end-to-end data management platform designed for IoT scenarios. The system requires fewer resources for a similar workload compared to large-scale stream processors such as Kafka [6] and Flink [1], making it suitable for onboard devices with limited CPU or memory [9]. By processing data at the edge, it reduces the reliance on strong or constant network connections. This approach lowers latency since events do not need to be sent to a cloud for processing. NebulaStream adapts to changing workloads [2], such as trains that may encounter areas of poor connectivity or abrupt changes in sensor outputs.

In addition, NebulaStream's execution engine offers a flexible plugin interface that enables the extension of different functionalities, allowing the integration of third-party libraries to support running complex queries for rich analysis [4].

### 2.2 MEOS

MEOS[2] (Mobility Engine Open Source) [11, 12] is a C library designed to manage both temporal and spatiotemporal data, focusing on mobility data. It incorporates data structures, such as spatiotemporal sequences and bounding boxes, based on the concepts of moving objects across temporal and spatial dimensions. Traditional solutions often separate location from time, complicating the tracking of moving objects. MEOS addresses this by storing and querying spatiotemporal data in a single framework, allowing tasks such as identifying a train's route through specific zones at precise time intervals. In addition, MEOS can be deployed within an edge computing environment, i.e., close to the data sources, such as a Raspberry Pi, and on cloud infrastructure, ensuring scalable data management. The library detaches storage from processing, enabling higher flexibility and efficiency when handling data.

---
[1]https://deck.gl
[2]https://libmeos.org/

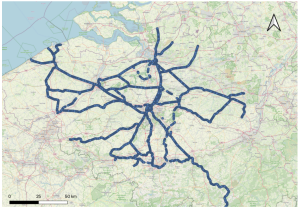

Figure 2: SNCB Data Visualization

## 2.3 NebulaMEOS

NebulaStream implements a plugin-based architecture that facilitates the integration of external components to extend its core functionality. In addition, the system features an expression framework, which allows the development of custom operators and functions through inheritance and composition. The framework also supports runtime operator definition through dynamic registration, enabling the integration of domain-specific operator logic, including calling MEOS functions.

NebulaMEOS adds custom operators, including `MeosAtStbox_Expression`, which incorporate spatial predicates such as `edwithin` and `tpoint_at_stbox`. The function `edwithin` checks if a geometry and a temporal point ever fall within a specified distance of each other, and the function `tpoint_at_stbox` returns a temporal point restricted to a spatiotemporal box. These predicates evaluate relationships in streaming data, allowing for spatiotemporal queries. These operators can, for example, filter event streams based on proximity constraints. MEOS extends the expressions processing framework to support tumbling, sliding, and threshold windows over spatiotemporal data streams. To this end, we extend the window definition expressions and operands. This feature enables the grouping and analysis of spatiotemporal data. Furthermore, we address the scalability demands in highly distributed IoT environments that require specialized spatiotemporal algorithms that can function on resource-constrained edge hardware. This entails developing lightweight data structures and operators capable of handling continuous data streams under strict latency and memory constraints.

## 3 Demonstration

We use a laptop with MacOS and a Raspberry Pi with Ubuntu to deploy NebulaMEOS for our demonstration.[3] We simulate the continuous event stream from a dataset originating from edge devices installed on six trains, with information such as GPS coordinates, battery voltage levels, and brake pressure readings over six months. This dataset offers information about the trains' operational and geographical position (Figure 2)

We define two categories of queries, Geofencing and Geospatial Complex Event Processing, to demonstrate how NebulaMEOS works to enable spatiotemporal queries in IoT environments. In addition, we report the ingestion rate and throughput per query.

---

[3]The demo video is at https://youtu.be/lQS11qhy7J0

### 3.1 Geofencing

A geofence is a boundary that limits a location. It can be created dynamically in a radius from the center of the area or by setting the boundaries to perimeters, such as the edges of a neighborhood. Geofencing supports analysis by establishing and supervising the pertinent geographical borders within the operation area. Our integration enables four use cases in geofencing. First, we implement location-based alert filtering. When the trains enter maintenance zones, the system will respond by discarding non-essential alerts such as speeding. Second, we perform location-based monitoring to track noise levels within specific regions and minimize the train's impact on neighborhoods. Third, the system suggests train speed based on GPS data, ensuring safe navigation through high-risk areas such as sharp curves or construction sites. Fourth, combining weather measurements will suggest safety measures in adverse conditions such as heavy snow and fog.

*Query 1: Location-Based Alert Filtering.* The system determines whether a train is within a maintenance area. Alerts such as speed violations or equipment malfunctions are filtered out if confirmed.

*Query 2: Location-Based Noise Monitoring.* The query monitors the sound levels outside the train. The system can relate some of these noise peaks to their geographical areas by bringing in a geospatial context. This spatial analysis allows for appropriate noise reduction in areas such as managing the engine output performance.

*Query 3: Dynamic Speed Limit.* By utilizing real-time GPS data, the system can enforce speed restrictions dynamically, adapting to specific zones, such as curves and other high-risk areas.

*Query 4: Weather-Based Speed Zones.* We integrate weather data from OpenMeteo[4] to suggest speed limits for zones with conditions such as heavy rain, snow, or fog, maintaining safety operations.

In Queries 1 to 4, we have achieved a throughput of 2.24 MB with 20K events per second (e/s).

### 3.2 Geospatial Complex Event Processing

Geospatial Complex Event Processing (GCEP) is a paradigm that enables detecting and analyzing patterns and relationships within spatiotemporal data streams. In the context of NebulaMEOS, GCEP facilitates the real-time processing of data generated by trains, enabling proactive decision-making. The GCEP is extended from the work presented in [10]. The system offers an alternative to existing GCEP systems, such as [5], by integrating real-time spatiotemporal analytics and pushing down computation to IoT devices. Our integration enables four use cases in GCEP. First, NebulaMEOS monitors temperature and battery usage when on battery power, keeping track of nearby workshops. Second, our system enables the detection of a heavy load of passengers and suggests adjusting temperature and lighting. Third, the NebulaMEOS system detects unscheduled stops outside stations and workshops. Fourth, the system monitors brake usage and brake pressures.

*Query 5: Battery Monitoring.* We ensure the battery's charge and discharge cycles follow a predefined curve. Deviations might indicate that the battery's health is decreasing. The system generates alerts

---

[4]https://open-meteo.com/

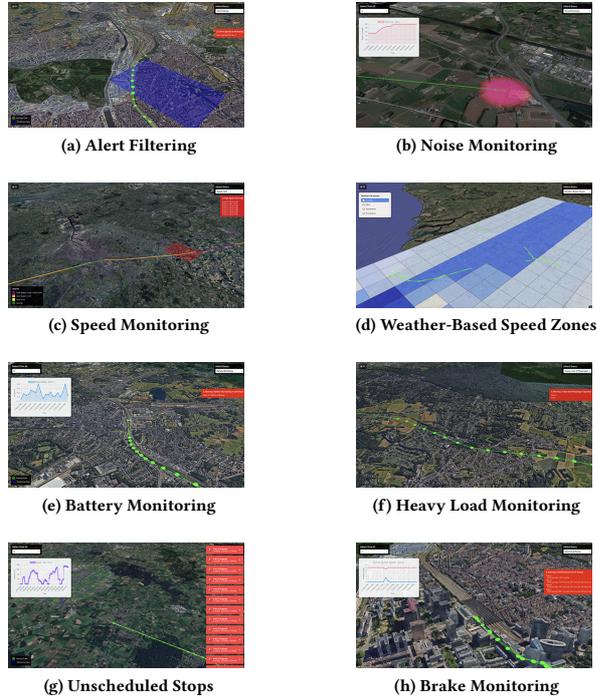

Figure 3: Queries' Visualizations

(a) Alert Filtering
(b) Noise Monitoring
(c) Speed Monitoring
(d) Weather-Based Speed Zones
(e) Battery Monitoring
(f) Heavy Load Monitoring
(g) Unscheduled Stops
(h) Brake Monitoring

for overheating and excessive discharge. In addition, it queries nearby workshops in case of emergencies. We have obtained a throughput of 0.61 MB with 8K e/s.

*Query 6: Heavy Passenger Load.* We estimate the number of passengers to determine whether there are no free seats to verify whether the number of trains put into service is sufficient for the number of passengers, such that an extra train can be added in the following days if that is not the case. We have obtained a throughput of 3.68 MB with 32K e/s.

*Query 7: Unscheduled Stops.* The stop is flagged as unscheduled if the train stops outside designated zones. This alert helps operators act, whether they need to send help or investigate mechanical problems. It also prevents unauthorized halts that could affect service reliability. We have a throughput of 0.40 MB with 10K e/s.

*Query 8: Monitoring Brakes.* The system detects patterns such as repeated emergency brakes in specific track segments or persistent low-pressure readings that could indicate a decrease in the brake's effectiveness. We achieved a throughput of 2.24 MB with 20K e/s.

### 3.3 Visualizations and User Interaction

The demonstration includes visualizations that allow users to assess the challenges of train operations and the analysis that can be performed with the NebulaMEOS system (Figure 3). The visualization was built in Deck.gl and uses Kafka to input data.

We show real data collected from trains, including coordinate positions, speeds, and brake pressure. Users can select a query and a train to see details about the route. Each query uses colors and labels that spotlight trends and help operators detect potential issues. The queries react with alerts and flags as the stream flows, producing a visual alert when the query condition is satisfied.

## 4 Conclusion

In this demonstration, we introduce NebulaMEOS, which combines MEOS, a spatiotemporal processing library, with NebulaStream, a scalable data management system designed for IoT applications. NebulaMEOS allows spatiotemporal functionalities that enable real-time processing and analysis of streaming data. To validate our approach, we conducted queries on data collected from edge devices installed on trains operated by SNCB.

For future work, we plan to enhance the spatiotemporal stream processing capabilities within NebulaMEOS by defining additional operations from MEOS. These will include the development of trajectory-based functions in addition to the point-based functions described in this demonstration. We also aim to define aggregation functions that can work with elements within the stream to answer queries such as identifying the top-k nearest trains. We will also deploy the system directly on the edge devices within the trains.

## Acknowledgments

This work was partially funded by the EU's Horizon Europe research program grant No. 101070279 MobiSpaces.